\def\ha{H$\alpha$~}

\def\sm{$\cal M_\odot$~}
\def\sma{$\cal M_\odot$}
\def\ma{$\cal M$}

\def\ss{\smallskip}
\def\ms{\medskip}

\documentclass{aa}
\usepackage{graphics,times}
\begin{document}

   \thesaurus{11.03.2,11.05.2,11.19.3,13.09.1}
   \title{LWS spectroscopy of the luminous Blue Compact Galaxy Haro 
11\thanks
{Based on observations with ISO, an ESA project with instruments
funded by ESA Member States (especially the PI countries: France, Germany,
the Netherlands and the United Kingdom) with the participation of ISAS and
NASA.}}

\markboth{Bergvall et al.:  ISO LWS spectroscopy of Haro 11}{}
\authorrunning{Bergvall et al.}
\titlerunning{ISO LWS observations of Haro 11}
\author{Nils Bergvall$^1$ \and Josefa Masegosa$^2$
\and
G\"oran \"Ostlin$^3$ \and Jose  Cernicharo$^4$}

\offprints{ N. Bergvall, email: nils.bergvall@astro.uu.se }

\institute{1) Astronomiska observatoriet, Box 515, S-75120 Uppsala, Sweden; 
\\
2) Instituto de Astrofisica de Andalucia, CSIC, Apdo. 3004, E-18080 Granada,
Spain; \\
3) Stockholm Observatory, SE-133 36 Saltsj\"obaden, Sweden; \\
4) Instituto de Estructura de la Materia, CSIC, Serrano 121, E-28006 Madrid,
Spain}
\date{Received 0000; accepted 0000}

   \maketitle

   \begin{abstract}

We present the results of far infrared (FIR) spectroscopy of the luminous 
blue
compact galaxy (BCG) \object{Haro 11}
(\object{ESO 350-IG38}) obtained with the
Long Wavelength Spectrometer (LWS) on the Infrared Space Observatory (ISO) 
in
low
resolution mode.
This metal poor dwarf merger is an extremely hot IRAS source. We discuss the
balance between dust and
line cooling in the photodissociated regions (PDR), in particular the role 
of
the [CII]$\lambda$158$\mu$
line, and derive the basic properties of the PDR gas and estimates of the 
gas
and
dust masses. The
mass of the PDRs,  2$^{+2}_{-1}$~10$^8$ \sma, is comparable to that of the
ionized gas and exceeds
the observed upper limit of the HI mass. The gas/dust mass ratio is low,
indicating that the galaxy
contains little cold dust. The low metallicity, the intense radiation field
and the low column density of Haro 11 results in an extremely high [CII]/CO 
flux
ratio and probably also a very high \ma (H$_2$)/L$_{CO}$ conversion factor.
Therefore CO is a poor indicator of the H$_2$ mass in starburst dwarf 
galaxies.

After a reanalysis we confirm the claimed correlation (Malhotra et al.
\cite{malhotra}) between the
[CII]158$\mu$/FIR flux ratio and the  IRAS f$_{60}$/f$_{100}$ dust 
temperature
and reduce the
scatter. We find that
Haro 11 deviates from the relationship being brighter in [CII] than what 
would
be
expected, if the
mechanism proposed by Malhotra et al. is dominant.
As an alternative (or complementary) explanation we propose that the
[CII]158$\mu$/FIR versus
f$_{60}$/f$_{100}$
relationship is caused by an increasing optical depth with increasing IRAS
temperature. The low
metallicity of Haro 11 and its extreme starburst properties probably allows 
the
medium to be thin despite
its high
f$_{60}$/f$_{100}$ ratio. This leaves room for a more optimistic view on
the
possibilities to detect
massive starforming mergers at high redshifts, using the [CII] line.

      \keywords{Galaxies: compact, Galaxies: evolution, Galaxies: starburst,
Galaxies: ISM, Galaxies: Haro 11, Galaxies: ESO 350-IG38,
Infrared: galaxies}
   \end{abstract}

%

\section{Introduction}

Blue compact galaxies (BCGs) are characterized by their high star formation
rate and
low chemical abundances. The high ratio between blue luminosity and HI mass,
L$_B$/\ma$_{HI}$, combined with a high relative HI mass fraction, indicates 
a
high star formation
efficiency and a high gas consumption rate.
The apparent conflict between intense star formation and low chemical
abundances can
only be explained if the gas
containing the recently produced elements is expelled from the galaxy, if
fresh gas is accreted onto
the galaxy or if
the galaxy is young (e.g. Searle \& Sargent \cite{searle}; Makarova \&
Karachentsev \cite{makarova}; Papaderos
et al. \cite{papaderos}).
We still have rather vague
ideas about what is triggering the intense global burst. Our previous
observations (\"Ostlin et al. \cite{ostlin1}, \cite{ostlin2}) indicate that
gas dynamical
instabilities
in connection with
mergers could be the key mechanism. Under such circumstances the heating
processes
of the ISM could become a complicated mixture of shocks caused by infalling
clouds and winds from massive young stars and on the other hand a strong
radiation
field from
the starburst.

While optical and radio observations give us important information about the
conditions in the hot ionized and the cold neutral gas in star forming 
galaxies
we often lack
information about the the ISM in the transition zone, the photodissociation
regions (PDRs). The gas
in PDRs contains neutral atoms and ions of low ionization
potential ($\chi <$ 13.6 eV), e.g. C$^+$. With the different possible 
heating
mechanisms involved one
would
expect that the geometry and spatial extension of the ISM and the PDRs would 
be
quite different  in galaxies like Haro 11 and those with quiet star 
formation
activity
or with active nuclei, resulting in radically different spectral
signatures.

Spectral diagnostics are valuable tools in dissecting the balance between
different heating and
cooling processes and the physical
conditions in the gas. The first observations of [OI]63$\mu$ in M82 was
presented
by Watson et al.
(\cite{watson2}). Crawford
et al. (\cite{crawford}) for the first time showed that important 
information
about the PDR physics in
external galaxies can be obtained from the collisionally excited
[CII]$\lambda$158$\mu$. Later,
several studies
of star forming regions in the Milky Way as well as other starburst galaxies 
and
ultraluminous
infrared galaxies ('ULIGs', possibly hiding active nuclei) have
demonstrated the power of more detailed far-IR spectroscopy (e.g. Colbert et 
al.
\cite{colbert}).
An advantage with observations in the IR
region is that we diminish the influence of dust extinction, allowing a more
direct
comparison between 'naked' starbursts in low mass galaxies and hidden 
starbursts
in
ULIGs.

The fact that [CII]$\lambda$158$\mu$ emission constitutes the peak of the 
flux
distribution f$_{\nu}$
in
starforming galaxies has led to a discussion about the possibility of using
[CII] to detect young galaxies at high redshifts (e.g. Petrosian et al.
\cite{petrosian} and the
detailed discussion by Stark \cite{stark}). Despite
some emerging pessimism about this possibility (Gerin \& Phillips 
\cite{gerin})
we will argue that this still is a viable idea.

As part of our multifrequency study of three of the most luminous BCGs
in the southern
hemisphere this report will focus on observations using the Long Wavelength
Spectrometer
(LWS) (Clegg et al. \cite{clegg}) on the
{\it Infrared Space Observatory} (ISO)
(Kessler et al. \cite{kessler}) of the luminous BCG Haro 11.
Only a few emission lines in the LWS window are detectable in most
nearby BCGs in a reasonably long integration time. In our case we chose to
obtain clear detection of a few of these lines, with a sufficiently high
S/N to
allow comparisons with models and other galaxies. These lines are the
[OI]63$\mu$, [OIII]88$\mu$ and [CII]158$\mu$ lines, which constitute 
important
coolants in the far-IR regions. Among these lines, [CII] is of particular
importance for reasons just mentioned and discussed below.
In addition to these three lines, we obtained an upper limit on the flux in 
the
[OI]145$\mu$ line in one of the adjoining detectors.

\section{Observations and reductions}

The observations were carried out during the routine observation phase of
ISO in
the low-resolution LWS02 observation mode with a resolution of 0.6 $\mu$.
We used a scan width of 7 resolution elements, 8 samples per element and 
fast
scanning.
Each spectrum was scanned 6-111 times with a nominal 0.50 s integration time
giving in total an effective integration time of 2820 s.

   \begin{figure*}
      \vspace{0cm}
	\resizebox{\hsize}{12cm}{\includegraphics{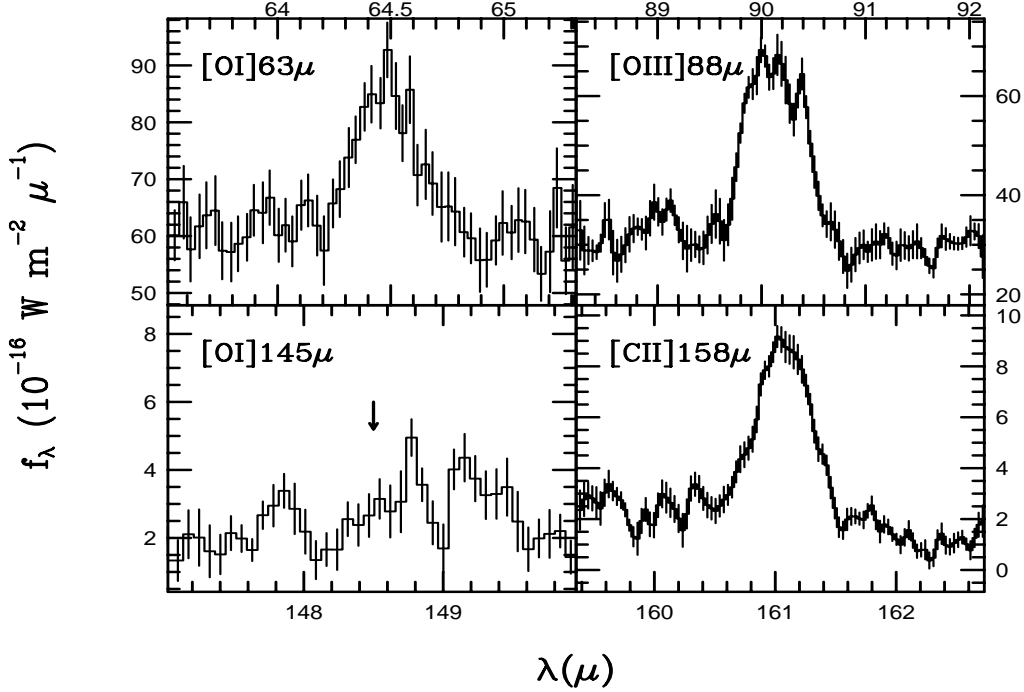}}
      \caption[]{The ISO LWS emission lines of Haro 11. 3 detections are
presented. An upper limit of the flux of the [OI]$\lambda$145$\mu$ line was
derived from the region at the expected position of the line, marked with an
arrow. The bars indicate mean errors per pixel.}
         \label{obs}
   \end{figure*}

The data presented in this paper have been taken from pipeline 6.1. All the glitches in each individual scan have been removed
by using the Interactive Analysis Software Package.
About 15\% of the total scans were rejected due to heavy
cosmic rays contamination. In addition to the requested data
the LWS spectrometer provided data on all the other detectors.
These data have been also analyzed and an upper limit on the
fine structure line of [OI]$\lambda$145$\mu$ has been derived.
The resulting data were resampled and coadded. The data are shown in
figure 1.

Exept for the LWS data we also present some additional previously 
unpublished
data in Table 1. The CO observations were carried out during June 9-12 1988
with the SEST telescope equipped with cooled Schottky receivers at 115 GHz
(J\"ors\"ater \& Bergvall \cite{steven}). Six blue
compact galaxies were observed and none was detected.

Optical spectroscopy was obtained with ESOs 1.5-m telescope in 1983, using
the Image Dissector Scanner and an aperture of 4"x4" arcseconds$^2$ 
(Bergvall
\& Olofsson \cite{bergvall1}). Additional data were obtained with the IDS at 
the
ESO 3.6-m telescope in 1986. Standard reduction techniques were used. More information is found in Bergvall \&  \"Osttlin (\cite{bergvall2}).

\section{Results}

\subsection{General properties of the target galaxy}

Haro 11 (ESO 338-IG04) is a galaxy with extreme starburst properties, summarized in Table
1.  In the following we will
assume a Hubble constant of H$_0$ = 65 km~s$^{-1}$Mpc$^{-1}$.
Most of the data are from our previous observations in the UV-radio region
(Bergvall
\& Olofsson \cite{bergvall1}; \"Ostlin et al \cite{ostlin1}, \cite{ostlin2};
Bergvall \& \"Ostlin \cite{bergvall2} and other unpublished data). Our
observations show three
conspicious
condensations in the centre embedded in a regular \ha halo. HST images 
(Malkan
et
al. \cite{malkan};
\"Ostlin et al \cite{ostlin2}) reveal a merger morphology. Faint whisps and
indications of shell structures characteristic of galactic
disk progenitors are seen in the outer regions. Our studies of the 
kinematics
of the
ionized gas indicate that the starburst is triggered by coalescing clouds
involved in the merger process. The \ha velocity field is requires a
mass of the main body of $>$10$^9$ \sma~(\"Ostlin et al. \cite{ostlin1};
\cite{ostlin2}).

The oxygen abundance in the central region was calculated from the optical
spectra, using the standard method. We assumed a homogeneous single
metallicity model with two temperature zones. The electron temperature of 
the
O$^{++}$ zone was derived from the [OIII]$\lambda$4363/([OIII]$\lambda
\lambda$(4959+5007) ratio and that of the O$^{+}$ zone from the 
semi-empirical
relationship derived by Vila-Costas and Edmunds (\cite{vilacostas}). The
density was derived from the [SII]$\lambda$6717/6731 line ratio and was
found to be low. With this approach we obtain a metallicity close to 10\% 
solar.
Such a value is unusually low for a galaxy
of this luminosity and deviates strongly (Bergvall et al. \cite{bergvall5}) from the metallicity-luminosity
relationship of normal dwarf galaxies (Skillman et al. \cite{skillman}). Part of the optical emission may have 
a
shock origin and we made an estimate how large the contribution from shocks 
may
be by mixing predicted line spectra from shocked regions (Raymond \cite{raymond}) with 
those
of HII regions from Stasi\'nska (\cite{stasinska1}). The best fits generally 
gave
a contribution from the shock component to H$\beta$ of $<$ 5\%. This
is in most cases not enough to change the result of the calculations of
the metallicities drastically.

A surprising fact is that no HI has been detected in the galaxy (Bergvall et
al. \cite{bergvall3}). With an upper limit of \ma $_{HI} = $10$^8$ \sm and
consequently \ma $_{HI}$/L$_{B,total}<$0.01, it seems to be remarkably
devoid of neutral hydrogen for its star formation rate.
One possible explanation is that the neutral gas is ionized by the central
starburst. Another possibility is that much of the gas is in molecular form.
The ISO observations discussed here, which relate to the transition zone
between the ionized and the cold molecular medium, will help to derive a
coherent picture of the different phases of the gas.

\begin{table}

\caption[]{General properties of Haro 11}
\begin{flushleft}
\begin{tabular} {lll}
\noalign{\ss}
\hline

& & Ref./Note \cr
\noalign{\smallskip}
\hline
\noalign{\smallskip}
$\alpha_{1950}$ & 00$^h$34$^m$25\fs 7 \cr
$\delta_{1950}$ & -33$^{\circ}$49'49"
\cr
m$_B $ & 14.3 & 1/a \cr
v$_0$ & 6175$\pm$10  km~s$^{-1}$ & 1/b \cr
r$$ & 92 Mpc & 3 \cr
M$_B$ & -20.5 & a\cr
a$_{\mu_B=26.5}$ & 9 kpc & 2/c\cr
B-V$$ & 0.5 & 1/a\cr
12+[O/H] & 7.9 & 2\cr
L$_{Ly\alpha}$ & 9~10$^{35}$ W & 3\cr
L$_{H_\alpha}$ & 3.2~10$^{35}$ W & 2\cr
L$_{FIR}$ & 7.4~10$^{10}$ L$_\odot$\cr
f$_{60}$/f$_{100}$ & 1.48 \cr
L$_{FIR}$/L$_B$ & 5.7 \cr
L$_{CO(1-0)}$ & $\leq$ 1.0~10$^{29}$ W & 4\cr
log(L$_{CO}$/L$_{FIR}$) & $\leq$-8.5 \cr
\ma $_{H_2}$ & $\leq$ 10$^8$ \sma & d\cr
\ma $_{HI}$ & $\leq$ 10$^8$ \sma & 4\cr
\ma $_{HI}$/L$_B$ & $\leq$ 0.01 \cr
\noalign{\smallskip}
\hline
\smallskip
\end{tabular}
\\

a) Corrected for galactic extinction (Burstein \& Heiles \cite{burstein}) \\
b) Radial velocity, corrected to the local group according to
v$_0$=v+300 sin$l$ cos$b$, where v is the heliocentric velocity and $l$ and
$b$ are the galactic coordinates \\
c) r: distance; a: major axis diameter \\
d) Using a conversion factor of  3~10$^{20}$ cm$^{-2}$ (K km s$^{-1}$) 
$^{-1}$
(Sanders et al. \cite{sanders}). \\
\ms
References: 1) Bergvall \& Olofsson \cite{bergvall1}, 2) Bergvall \& 
\"Ostlin
\cite{bergvall2}, 3) Bergvall \& Olofsson, \cite{bergvall4}, 4) J\"ors\"ater 
\&
Bergvall, \cite{steven}\\
\end{flushleft}
\end{table}

\subsection{The dust properties}

From the 60 and 100 $\mu$ IRAS fluxes and assuming a FIR grain opacity of
$\kappa$ =
2.5~10$^3\lambda^{-1}$ cm$^2$g$^{-1}$ (Hildebrand \cite{hildebrand}), where
$\lambda$ is the
wavelength in microns, we calculate a dust temperature of T$_{dust}$ = 49K. 
We
can then obtain a
rough estimate of the mass of the {\it warm} dust from (Devereux \& Young
\cite{devereux})

$$ {\cal M}_{dust} = 4.58~f_{100}r^2(e^{144/T_{dust}}-1)~{\cal M}_\odot$$

f$_{100}$ is in Jy and r is the distance in Mpc. We obtain ${\cal
M}_{dust}\sim$
4~10$^6$ \sma.

It is interesting to compare gas-to-dust mass ratio with that of other gas 
rich
galaxies. In Sect.
3.4.3 we find that the total hydrogen gas mass, i.e. HI, HII and H$_2$, 
is
approximately
10$^9$ \sma. We thus obtain ${\cal M}_{gas}$/${\cal M}_{dust}\sim$ 200.
In a study of normal spiral galaxies, Devereux
\& Young
(\cite{devereux}), taking into account only the inner part of the HI disk
containing the warm dust,
found that ${\cal M}_{gas}$/${\cal M}_{dust}\sim$ 1100. There is thus a difference
a factor of 5-6 difference between Haro 11 and normal spirals. If the dust 
content
correlates with metallicity, one would expect that  ${\cal M}_{gas}$/${\cal
M}_{dust}$ would
increase with decreasing metallicity and indeed this is the case for dIrrs 
as was shown by Lisenfeld
\& Ferrara (\cite{lisenfeld}). If we apply this metallicity correction in the comparison, it would lead to an even larger
difference between Haro 11 and the spirals in the  ${\cal M}_{gas}$/${\cal M}_{dust}$ ratio. The high value for spiral galaxies has been
shown (e.g. Chini et al. \cite{chini}) to be due to the fact that a large 
amount of the dust is of the cold, 'cirrus' type, radiating at longer wavelengths and thus not contributing to the mass estimates based on the IRAS data. The fact that ${\cal M}_{gas}$/${\cal M}_{dust}$ of Haro 11 is so low (in fact close to the value found for our own Galaxy, including the {\it total dust mass}, i.e. 
${\cal M}_{gas}$/${\cal M}_{dust,total}\sim$ 100-150) indicates that the
galaxy contains very little dust at low temperature.

\subsection{The LWS spectra}

Figure 1 shows the spectral profiles of the three detected lines and the 
region
around [OI]145$\mu$, for which we derived an upper limit. Table 2
summarizes the spectral data. From our optical spectra we have derived an
extinction coefficient in the brightest region of the galaxy of A$_V 
\approx$
0.8 mag. Thus the corrections for galactic extinction in the  infrared are
considerably less
than the observational errors in the observed emission lines (Adams et al.
\cite{adams}).

\begin{table}

\caption[]{Observed LWS data. The table shows the measured line flux, its
estimated mean error and equivalent width (W) and the adjacent continuum 
level
flux
density.}
\begin{flushleft}
\begin{tabular} {llll}
\noalign{\smallskip}
\hline
\noalign{\smallskip}
Line id. & Flux & W & Continuum flux density  \\
& (10$^{-16}$W~m$^{-2}$) &   ($\mu$) & (10$^{-16}$W~m$^{-2}~\mu^{-1}$)  \\
\noalign{\smallskip}
\hline
\noalign{\smallskip}
[OI]63$\mu$ & 9.2 $\pm$ 1.8 & .22 & 42 \cr
[OIII]88$\mu$ & 25 $\pm$ 0.8 & .84 & 33 \cr
[OI]145$\mu$ & $\leq$ 0.3 & $\leq$ 0.2 & 1.7 \cr
[CII]158$\mu$ & 3.9 $\pm$ 0.1 & 2.0 & 2.1 \cr
\noalign{\smallskip}
\hline
\end{tabular}
\end{flushleft}
\end{table}

\subsection{Properties of the PDR gas}

One of the purposes of this study is to investigate how much the PDR gas
contributes to the total mass in this type of objects, and in particular
Haro 11 due to its remarkably low HI content. It is known that
the relative volume of the PDR that occupies the weakly ionized (only low
ionization stages) gas may be quite substantial. Regions having a total
extinction of A$_V\leq$10 are partly photodissociated (see e.g. Crawford et 
al.
\cite{crawford} or Tielens \& Hollenbach
\cite{tielens} for a general discussion about the PDR physics). This is the 
situation
for most of the neutral gas in the Galaxy. Estimates also show that $\leq$ 
40\%
of the total gas mass in starburst nuclei are PDRs, i.e. the 
photodissociated gas
is the dominant phase in these extreme environments (Stacey et al. 
\cite{stacey}).
The most widely used model for calculating the energy balance and obtain
predictions about emissivities of the gas and dust was presented by Tielens
\& Hollenbach (\cite{tielens}). Later upgrades and applications have been
discussed by Hollenbach et al. (\cite{hollenbach}), Wolfire et al.
(\cite{wolfire}) and Kaufman et al. (\cite{kaufman}).

The Kaufman et al. standard model is
based on an oxygen abundance of 40\% solar, close to the local Galactic value and a factor of 4-5 higher than in Haro11.
From our derived carbon abundance in ESO 338-IG04 (Tol 1924-416), a galaxy
quite similar to Haro 11, we may assume that N(C)/N(O) $\approx$ 0.2 
(Bergvall
\cite{bergvall}). Similar values have been obtained for other metal 
poor
galaxies (Garnett et al. \cite{garnett}). Thus, oxygen and carbon are underabundant with factors 4-8
with respect to the standard setup in Kaufmans et al. models  (their models
have N(C)/N(O) $\approx$ 0.5). The predicted temperature of the surface
gas layers obtained from their basic diagnostic diagrams is therefore 
probably
somewhat overestimated while still the line
intensities are reasonably correct (Wolfire et al. \cite{wolfire}). However,
Kaufman et al. also discuss a case of lower metallicity, about 4\% solar and
present diagrams showing how [CII], FIR and CO luminosities relate to 
density, n,
incident far UV radiation intensity, G$_0$, and the two metallicities $\approx$
4 and 40\% solar. In the following we use these models to estimate
some fundamental parameters of the PDR regions.

\subsubsection{Optical thickness}

To enable a comparison of the line intensities with the models of Kaufman et
al., we first estimate the contribution from HII regions to the line 
intensities
and also the optical depth of the lines. The optical emission lines 
(Bergvall
\& Olofsson \cite{bergvall1} and newer data) have been used
together with  recent photoionization models (Stasi\'nska \cite{stasinska}, 
and
updates) to estimate the contribution to the far-IR lines from the HII 
regions.
We used our observed line intensities in [OI], [OII], [OIII], [NII], [SII], 
HeI and
HeII, corrected for extinction, underlying absorption and shock 
contribution to select the best fitting HII region models from
Stasi\'nskas library. As a result we find that the measured intensity  of 
the
[OIII]$\lambda$88$\mu$ line is fully consistent with photoionization by a
cluster of stars with T$_{eff}$ of the order of 35000-40000 K. For the other
3 lines observed with the LWS,
the contribution from HII regions to the measured intensity is strongest
in [CII] but is still $\leq$ 20\%. Thus, when discussing the [CII] line flux
and flux ratios, we can directly compare to the predictions from Kaufman et 
al..

In a study
of two star forming complexes in the Large Magellanic Cloud, Israel et al.
(\cite{israel}) concluded that these regions were optically thin at a
{\it minimum} column density of 3~10$^{21}$ cm$^{-2}$. Since these cloud
aggregates probably have similar properties as Haro 11 we can conclude that
the star forming regions in Haro 11 on  a local scale probably are optically
thin in 158$\mu$. The optical extinction derived from the
H$\alpha$/H$\beta$-ratio in Haro 11 is only A$_V$ = 0.8 so we don't expect 
that
the [CII] line is optically thick even on a global scale. But the derived
extinction depends on the spatial distribution of the HII regions so to 
assure
that we don't have conditions that allow hidden sources in the central area 
we
will estimate the optical depth in [CII], $\tau_{CII}$. This calculation 
will
also be used to make estimates of optical depths of the other galaxies 
involved
in the discussion.

In the high-temperature (T$\gg$92K, the excitation potential for [CII]),
high-density limit (n$\gg$4~10$^3$ cm$^{-3}$, the limit for collisional
deexcitation by H and H$_2$),
$\tau_{[CII]}$ can be expressed as (Crawford et al. \cite{crawford})

$$ \tau_{[CII]} = 2.3~10^{-16}{{N_{C^+}}\over {T_{C+}\sigma _v}} $$

where N$_{C+}$ is the column density of C$^+$ in cm$^{-2}$, T$_{C+}$ is the
temperature of the gas in
K and $\sigma _v$ is the velocity dispersion in km~s$^{-1}$.  If we as a 
first
rough approximation assume that the global distribution of the gas is close
to a homogeneous exponential disk we can derive the central column density 
in
atoms cm$^{-2}$

$$ N_{gas} = 1.3~10^{14}{{\cal M}_{gas}\over {2 \pi h_{gas}^2}} $$

where $\cal M$$_{gas}$ is the mass of the gas in solar masses and h$_{gas}$ 
is
the scale length in
kpc. As a first approximation we will assume h$_{C^+}$=h$_{H\alpha}$. From 
our
H$\alpha$ images we
derive h$_{H\alpha}\sim$ 2" = 1 kpc. As an estimate of the gas mass we will 
use
our upper observational limit of
${\cal M}_{HI}$, i.e. around 10$^8$ \sma. With this we obtain a central 
column
density in HI of N$_{HI}\sim$ 2~10$^{21}$ atoms cm$^{-2}$. Later
we will find it
consistent to assume that most of the HI actually is associated with the PDR
regions. Part of the C$^+$ region in the PDR also contains molecular gas. In
sect. 3.4.2 we will show that this may increase the estimated column density
with a factor of 2.

The C$^+$ column density can now be derived from

$$ N_{C^+} \approx 1.3~10^{14} X(C^+){{{\cal M}_{HI}}\over {2 \pi
h_{H\alpha}^2}} $$

where X(C$^+$) is the relative abundance of carbon ions to hydrogen atoms. 
With
these approximation
we can write the optical depth as

$$  \tau_{[CII]} \leq 0.03 X(C^+){{{\cal M}_{HI}}\over {T_{C+}\sigma _v ~ 2 
\pi
h_{H\alpha}^2}} $$

Since most of the carbon in the [CII] emitting region is in ionized form we 
will
assume
X(C$^+$)$\approx$X(C$_{total}$)$\approx{N(C)\over N(O)}$~X(O)$\approx$
1.2~10$^{-5}$. In starburst
galaxies T$_{C+}$ is typically a few hundred K. We will later iteratively 
derive
T$_{C+}$ and obtain
a value below 400K. From our Fabry-Perot observations in H$\alpha$ of the
central
region we estimate
$\sigma _v \sim$ 60 km s$^{-1}$. We then obtain $\tau_{[CII]} \leq$ 0.001. 
In
order to derive a more
realistic estimate, we have to take into account that the gas has a lumpy,
fractal distribution. But
it is sufficient to note that, assuming a volume filling factor $\alpha$ of 
a
few
per mille (Crawford
et al. \cite{crawford}), the surface density would increase with $\sim 
\alpha
^{-2/3}$, i.e. a few tens.  Thus it still would be
safe to assume that the medium is thin in [CII]. The optically thick
limit at a chemical abundance similar to Haro 11 would be at a mean column
density of N$_{HI}\sim$
10$^{23}$ atoms cm$^{-2}$ and with solar abundances N$_{HI}\sim$ 10$^{22}$
atoms cm$^{-2}$, still with the assumption that the filling factor is a few 
per
mille.

\subsubsection{The mass of the PDR gas}

From the diagnostic diagrams by Kaufman et al. we can obtain the
density and intensity of the incident far-UV flux G$_0$ in the PDR region.
We use two diagrams for this purpose, one utilizing the
[OI]63$\mu$/[CII]158$\mu$ intensity ratio and one utilizing the
([OI]63$\mu$+[CII]158$\mu$)/FIR ratio. As mentioned above, before
we use the predictions we need to correct for the difference in
metallicity between Haro 11 and the models, a factor of 3-4. We can
estimate what effect this may have by comparing the predicted fluxes
of these lines in an HII region at the different metallicities. The column
densities and thus the volume emissivities of the low ionization
lines are lower in HII regions than in a typical PDR region but the
flux ratios should vary in approximately the same way. From Stasi\'nskas
models we find that the difference in flux ratios, assuming a constant 
N(C)/N(O), is
about 2 \%. N(C)/N(O) probably differ with a factor of about 2 however,
and so we should make an appropriate correction for this which means
a factor of 1.5-2 in the [OI]63$\mu$/[CII]158$\mu$ ratio. Kaufman et al.
show in later diagrams that the metallicity has a minor effect on the
[CII]158$\mu$/FIR ratio in the region of the parameter space where
we are residing.

We now feel confident to compare the slightly corrected line intensities 
with the predictions from the PDR models. Using the [OI]63$\mu$, [CII]158$\mu$ line fluxes and the FIR flux, Kaufmans et al. diagnostic diagrams show that G$_0$ is 1-2~10$^3$ in units of the Galactic far-UV
field (1.6~10$^{-6}$~W~m$^{-2}$). The obtained value for the density is
n $\approx$ 2~10$^3$ cm$^{-3}$ and thus G$_0$/n $\approx$ 1-2 cm$^3$.
These are typical values for normal starforming regions and galaxies 
(Tielens \& Hollenbach, \cite{tielens}; Wolfire et al. \cite{wolfire}; Carral 
et al. \cite{carral}; Fischer et al. \cite{fischer}). The surface temperature 
is T $\sim$ 400K. The data are summarized in Table 3. With
these data at hand we can obtain the predicted [OI]145$\mu$/[OI]63$\mu$ 
ratio which is approximately 0.06. This is within the errors of the
measured line ratio and thus supports the reliability of the model.

From the measured [CII] flux we may now calculate the mass of the warm
atomic gas according to (e.g. Wolfire et al. \cite{wolfire})

$$ {\cal M}_{PDR} = 5.96~10^{15}{{r^2~f_{[CII]}~m_H}\over {\Lambda
(CII)~X(C^+)}} \cal M_\odot $$

where r is the distance to the galaxy in Mpc,  f$_{[CII]}$ is the [CII] flux 
in W m$^{-2}$, m$_H$ is the mass of the hydrogen atom in kg, and  $\Lambda$(CII) is the cooling rate per atom in the 158$\mu$ line in W. In the high-temperature 
and high-density limit $\Lambda$(CII) will be nearly independent
of the temperature (Wolfire et al. \cite{wolfire}). $\Lambda$(CII) can
therefore be rather safely estimated and we obtain $\Lambda$(CII) $\approx$
1.3~10$^{-26}$ W atom$^{-1}$. In the surface layers of the PDR carbon is mainly 
in the low ionization monoatomic state so X(C$^+$)$\approx$ X(C).  We then 
derive \ma $_{PDR}\approx$ 2$^{+2}_{-1}$~10$^8$ \sma.

\subsubsection{The mass fractions of cold, warm and ionized gas}

The value we have derived for the mass of the PDR:s is larger than or 
comparable
to our upper
limit of the HI mass,
i.e. 10$^8$\sma, indicating that most of the HI gas is located in the PDR
regions
if these are
dominated by the atomic gas component. The total molecular gas mass in 
normal
star forming regions is
normally 5-10 times larger than in the PDR
regions. Data from
other starburst galaxies however indicate that considerably lower values are 
more
likely (e.g. Wolfire et al. \cite{wolfire}; Stacey et al. \cite{stacey}). 
This
is probably a metallicity effect. In
galaxies with low metallicities the UV radiation can penetrate deeper into 
the
molecular clouds, increasing the PDR zone relative to the CO core. An upper
limit is therefore probably around 10$^9$ \sm in our case. This mass 
estimate
may
be compared to the upper limit of the H$_2$ mass, derived from the CO (1-0)
observations which is $\leq$ 10$^8$ \sma. The calculation of this last value 
is
however rather uncertain due to the uncertainty in the conversion factor 
between
the CO flux and \ma $_{H_2}$ (Maloney \& Black \cite{maloney}, Taylor et al
\cite{taylor}). As just mentioned, the low metallicity also allows
a larger proportion of H$_2$ to be hidden in the CO dissociated zone. 
Therefore,
in addition to the problem caused by the uncertainty in the conversion 
factor
between CO and H$_2$, the CO fluxes are poor indicators of the H$_2$ masses 
in
low metallicity galaxies.

Using the available photometry of the halo and
the burst, our spectral evolutionary models predict a total stellar mass of 
1.5
10$^{10}$
\sm , while the observed rotation results in a mass of $\sim 2~10^9$ \sm,
assuming dynamical equilibrium (\"Ostlin et al. \cite{ostlin1};
\cite{ostlin2}). The discrepancy between the photometric and
dynamical
mass estimates can be resolved if the galaxy is dominated by velocity
dispersion,
which is reasonable if the width of the H$\alpha$ line reflects potential
motions,
or if the galaxy is not in dynamical equilibrium.

From the global H$\alpha$ luminosity and the mean density of the gas, 
derived
from our spectra, the mass of the ionized gas is estimated to be $\sim$
10$^8$ - 10$^9$ \sm
(\"Ostlin et al. \cite{ostlin1},\cite{ostlin2}). It is a remarkable fact
that the HI mass is so
low and also lower than or comparable to the estimated mass of ionized or
molecular gas. A possible explanation may be that HI to a
large extent has become molecular and/or ionized in the merging process. If 
this
is
correct, then the gas/total mass estimates may have been severely 
underestimated
in many dwarf starburst cases. This will complicate the efforts to 
understand
dwarf galaxy evolution in general. A summary of the mass estimates
is found in Table 4.

\begin{table}

\caption[]{Global data, based on the LWS observations}
\begin{flushleft}
\begin{tabular} {ll}
\noalign{\smallskip}
\hline
\noalign{\smallskip}
G$_0$ & 1-2~10$^3$ \cr
n & 2~10$^3$~cm$^{-3}$ \cr
G$_0$/n & $\approx$ 1 cm$^3$\cr
T & $<$500K \cr
L$_{[CII]158\mu}$/L$_{CO}$ & $\geq$4~10$^5$ \cr
L$_{[CII]158\mu}$/L$_{FIR}$ & 1.4~10$^{-3}$ \cr
(L$_{[OI]63\mu}$+L$_{[CII]158\mu}$)/FIR & 4.6~10$^{-3}$ \cr
L$_{[OI]145\mu}$/L$_{[OI]63\mu}$ & $\leq$ .03 \cr
L$_{[OI]158\mu}$/L$_{[OI]63\mu}$ & 0.42 \cr
\noalign{\smallskip}
\hline
\end{tabular}
\end{flushleft}
\end{table}

\begin{table}

\caption[]{Mass estimates (\sm)}
\begin{flushleft}
\begin{tabular} {ll}
\noalign{\smallskip}
\hline
\noalign{\smallskip}
HI & $\leq$ 10$^8$ \cr
HII & 10$^8$-10$^9$ \cr
H$_2$  & 10$^8$-10$^9$ \cr
PDR & 2$^{+2}_{-1}$~10$^8$ \cr
Warm dust & 4$^{+4}_{-2}$~10$^6$ \cr
Photometric & 1.5$^{+2}_{-0.5}$ 10$^{10}$ \cr
Rotational$^1$ & 2$^{+8}_{-1}$ 10$^9$ \cr
\noalign{\smallskip}
\hline
\end{tabular}

1) Assuming dynamical equilibrium

\end{flushleft}
\end{table}

   \begin{figure}
      \vspace{0cm}
	\resizebox{\hsize}{!}{\includegraphics{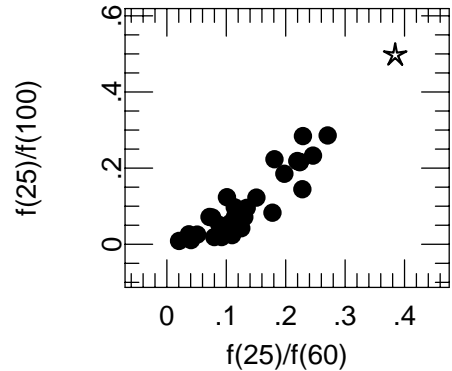}}
      \caption[]{The IRAS f$_{25}$/f$_{60}$ versus the f$_{25}$/f$_{100}$ 
micron
flux ratios for Haro
11 (star) and the comparison sample (filled dots)}
         \label{iras}
   \end{figure}

\subsection{Thermal balance and the [CII]/FIR ratio}

The important coolant [CII]$\lambda$158$\mu$ originates from the surface 
layers of PDRs where the C$^+$ ions are excited by photoelectrons ejected from 
grains heated by the UV radiation (Watson \cite{watson}). PAHs are believed to 
be the dominating agent of this process. Stacey et al. (\cite{stacey}) studied a mixed sample of spiral galaxies, starburst galaxies, giant molecular clouds and 
galactic star forming regions and noticed that while there is a linear relation
between the energy density of the UV field and the cooling rate of the dust, the
efficiency with which the [CII] line is excited seems to decrease with the 
intensity of the UV field. One should have in mind however that this is not a 
homogeneous sample. Malhotra et al. (\cite{malhotra}) investigated the same
relationship between the [CII] cooling and dust cooling for different types of 
galaxies, using the [CII]/FIR luminosity ratio. An anticorrelation was found 
between the [CII]/FIR ratio and dust temperature, as measured by the IRAS
f$_{60}$/f$_{100}$ ratio. After a discussion of several possible explanations,
Malhotra et al. found it most probable that this is a consequence of a 
decreasing efficiency of ejection of photoelectrons from the grains, reducing
the [CII] intensity. Theoretically it can be understood if the ratio of UV 
flux to gas density, and thus the dust temperature, becomes high because the 
dust grains would be more positively charged, forming a stronger Coloumb barrier against photoejection.

One should be aware, both as regards the results from Malhotra et al. and in 
the
following
discussion, that the sample for which useful data are available is quite
heterogeneous. The galaxy
types involved range from normal galaxies to Seyferts for which the relative
importance of different
heating sources (stars and AGNs) and different chemical abundances are not
satisfactorily controlled.
Moreover, in the first approximation, two dust components, warm and cold, 
having
different spatial
distribution, can be separated. This increases the complexity in the
interpretation of the global
data and the 60/100 index as well as the L$_{FIR}$/L$_B$ index cannot be
regarded
as clearcut tracers
of the star formation activity. As a consequence we expect that the scatter 
in
the discovered trends
will remain quite large.

We will now compare the data for Haro 11 with a data set similar to that 
used by
Malhotra et al. but
including more detected galaxies at relatively high
f$_{60}$/f$_{100}$. These data were obtained from Stacey et al.
(\cite{stacey}),
Luhman et al.
(\cite{luhman}), Malhotra et al. (\cite{malhotra}) and Lord at al.
(\cite{lord}).

For the most nearby
galaxies, having velocities $\leq$ 500 kms$^{-1}$, we have scanned the
literature
to find the most recent distance determinations derived from photometric
distance
indicators. This concerns
the galaxies M82, NGC 3109 (Kennicutt et al. \cite{kennicutt2}), NGC 4414
(Turner et al. \cite{turner}), NGC 4736 (Garman \& Young \cite{garman}), NGC
5128 (Tonry et al. \cite{tonry}), NGC 5194 (Sandage \cite{sandage}), NGC 
6946
(Freedman et al. \cite{freedman}) and IC 342 (Krismer et al.
\cite{krismer}). In order to take proper account of the mid-IR contribution 
to
the FIR luminosities, the FIR luminosities were calculated from FIR =
5.35~10$^5$~r$^2$(12.66~f$_{12}$+5.00~f$_{25}$+2.55~f$_{60}$+1.01~f$_{100}$)
L$_\odot$, where r is
the distance to the
galaxies in Mpc and f$_{12}$, f$_{25}$, f$_{60}$ and f$_{100}$ are the 
apparent
flux densities in
Jy (Belfort et al. \cite{belfort}), instead of the normal approximation, 
e.g.
FIR
=
3.75~10$^5$~r$^2$(2.58~f$_{60}$+1.0~f$_{100}$) L$_\odot$ (Lonsdale \& Helou
\cite{lonsdale}). This
choice makes a small but noticeable change in the distribution of the data 
in
the
diagrams.

Fig 2 displays the distribution of Haro 11 and the comparison sample in the 
IRAS
f$_{25}$/f$_{60}$
vs. f$_{25}$/f$_{100}$ diagram, showing the extreme position of Haro 11. 
Fig. 3
shows the FIR/B-60/100 diagram for the sample galaxies. In the diagram are 
also
indicated the loci of normal spiral galaxies and blue compact galaxies
(Dultzin
et al. \cite{deborah1}). As we see, Haro 11 also in this case lies at the
extreme
of the BCG distribution. The
extremely strong 25$\mu$ emission confirms the lack of cold dust and 
indicates
the presence of a strong nuclear starburst
(Dultzin-Hacyan et al. \cite{deborah1}; Hawarden et al \cite{hawarden}; see
also
Boulanger et al.
\cite{boulanger}; Taniguchi et al. \cite{taniguchi}), destroying the 
smallest
dust grains (e.g.
Desert \cite{desert}).

Fig. 4 and 5 correspond to Malhotras et al. diagram 1.
To extract the most important global parameters involved in the relation
reflected in fig. 4, we made
a cross
correlation test between the [CII]/FIR flux ratio, the IRAS data (i.e. the
ratios
between the 12, 25, 60 and 100$\mu$ fluxes), the FIR/B flux ratios and the 
FIR
flux. We found strong
correlations ($\geq 6 \sigma$) between only three of these parameters:
f$_{60}$/f$_{100}$,
FIR/B and the [CII]/FIR ratio.

In figure 4 the trend found by Malhotra et al. is confirmed and appears even
stronger. But it is also evident that the position of Haro 11 deviates from 
the
envelope defined by the other galaxies. The possible explanations for the
deviant
position of Haro 11 are either that there is an excess in the [CII] over FIR
luminosity or that the
f$_{60}$/f$_{100}$ temperature
is high, or both. Let us look at these two options one at a time. A consequence of lower metallicity is that the UV photons penetrate deeper into the cloud, thus increasing the volume of the PDR region at the expense of the colder molecular core. Strong observational support of this in terms of a high [CII]/CO flux ratio have been reported by e.g. Mochizuki et el. (\cite{mochizuki}), Poglitsch et al. (\cite{poglitsch}), Madden et al. (\cite{madden1}), Smith \& Madden (\cite{smith}) and Israel et al. (\cite{israel}). Haro 11 behaves in the same way with its extremely high [CII]/CO ratio ($>$ 4~10$^5$). The interesting
question is how much the [CII]/FIR ratio will be influenced by varying metallicity under these conditions. The models by Kaufman et al. show that [CII]/FIR is almost unaffected by metallicity variations in cases where G$_0$/n have normal values. This is also in agreement with observations. In 30 Dor and the cases discussed by Israel et al. however, the G$_0$/n ratio is low ($\leq$ 0.1) which results in a diluted radiation field, an increased heating efficiency and brighter [CII] emission. In Haro 11 the G$_0$/n ratio is normal (G$_0$/n=1 cm$^3$), as seen from table 3. So, both observations and modelling argue against an increase in [CII]/FIR in Haro 11 due to low metallicity. There is also another reason.
In fig. 5 Haro 11 adheres to the other galaxies in
the diagram. Therefore as a first guess it is reasonable to assume that
[CII]/FIR
is normal while
f$_{60}$/f$_{100}$ is deviating. 

We concluded in section 3.2 that Haro 11
contains very little warm or cold dust. Haro 11 is a merging system of galaxies and it is possible that the merging galaxies have lost their cold dust
components, e.g. extended disks, in the merger process. Is it possible that 
the lack of a 'normal' cold dust component could explain the high f$_{60}$/f$_{100}$ ratio? Would the contribution from such a component be sufficient to move the position of Haro 11 into the region of the other galaxies? It is an important question to ask since it separates global phenomena from local ones. We have made a rough
check of this possibility by selecting the galaxies in the comparison sample
having apparent sizes sufficiently large to allow us to obtain IRAS data
both from the very centre and also the integrated IRAS fluxes of the entire
galaxy. In this way we can have an idea about how much the extended component
influences the 60/100 ratio. It turns out that in the majority of the cases the 
IRAS temperature actually {\it increases} when the extended component is
included. Moreover, for the hottest cases, the hot component is always
significantly stronger than the warm/cold component, if existing. Our 
impression is therefore that the IRAS temperature of the warm dust of the most active regions in Haro 11 and the galaxies below Haro 11 in fig. 4 are similar.

Although we expect a large scatter in fig. 4 we should examine if there may 
be other parameters involved ruling the balance between [CII] and FIR emission and explaining part of the scatter in the diagram and the deviant position of Haro 11.  Since both the spectrum of the radiation field and the spatial distribution of gas and dust in general differs between low-mass starbursts and galaxies with AGNs, we could suspect that we would also find systematic differences in the [CII]/FIR ratios between these types of galaxies. When comparing galaxies classified in the optical region as Sy1, Sy2, liners and HII
galaxies we find that the HII galaxies appear to have a broad distribution in
f$_{60}$/f$_{100}$ while the Sy2 and liners tend to concentrate towards high IRAS temperatures and high FIR/B. Similarly, in a statistical investigation of the IR/optical properties of BCGs as related to Seyfert and Liner galaxies (Dultzin-Hacyan et al. \cite{deborah1}, \cite{deborah2}) it was shown that the f$_{25}$/f$_{100}$ is a powerful index that can be used to discriminate between pure starbursts and cases where AGNs are important for the radiation field.

\subsection{What rules the [CII]/FIR ratio?}

Several possible physical mechanisms have been proposed to explain the trends in [CII]/FIR seen in fig. 4 and 5 (e.g. Malhotra et al. \cite{malhotra}; Luhman et al. \cite{luhman}). As we already mentioned, Malhotra et al. argued that the reduced efficiency in photoejection due to a hard radiation field could be a plausible explanation. The data for Haro 11 however, acts as a counterexample of this. Both optical spectroscopy and the 'hot' IRAS indices indicate a hard
radiation field. Still, the [CII]/FIR is close to the values of normal, more
passive galaxies. An alternative explanation, that have been discussed by Malhotra et al. and others is that [CII] actually is optically thick in
many of the cases where the [CII]/FIR ratio is low. The general opinion seems to
be that the [CII] line is optically thin. These conclusions are based on estimates of the optical depth in a few well known cases, e.g. M82 (Crawford \cite{crawford}). But M82 is behaving well in the diagrams (see fig. 5)
and the central column density of the gas is not abnormal. From the discussion
above we see that, depending on the velocity field and the relative amount of molecular gas in the C$+$ zone, the medium could become optically thick in [CII] at a column density of N $\sim$ 10$^{22-23}$ cm$^{-2}$. Such high numbers can be
found in the central regions of some luminous starburst galaxies. Gerin \& Phillips (\cite{gerin}) elaborated on the low [CII]/FIR ratio in Arp 220 and it is interesting to compare this galaxy with M82 and Haro 11 (see fig. 5). They came to the conclusion that with an estimated column density of N $\sim$ 10$^{24}$ cm$^{-2}$, the {\it dust opacity and emission} would be sufficiently high to explain at least part of the low value of [CII]/FIR. But in fact there may be three sources responsible for reducing the ratio: dust opacity, dust emission and self absorption. What is maybe most important is how to
explain why the [CII]/L$_{Bol}$ ratio decreases with FIR/B. Therefore the combined effect of dust absorption and [CII] selfabsorption should
be investigated.

Fig. 6 shows the same data as in fig. 5, divided into different galaxy types,
Sy1, Sy2, HII and Liners. As we can see, there is a predominance of Sy2 galaxies
and Liners at high FIR/B. Most of these galaxies are ULIGs. There are strong indications that Liners are galaxies with massive nuclear starbursts and likewise some of the Sy2 galaxies. Many of these galaxies are probably remnants of massive merges. Even though the luminosity of some of them are not extreme, high column densities could possibly be obtained also in less massive nuclear starbursts, provided they are sufficiently compact. Such very compact nuclear starbursts are preferentially found in the type of galaxy we find at high FIR/B in the diagram (Heisler \& Vader \cite{heisler}). In a simplified form fig. 6
shows the effect of increasing optical depth on the sample. Where the gas is optically thick in [CII] the ratio between the [CII]/B flux ratio should be approximately constant. Where it is thin, [CII]/FIR should be constant.

   \begin{figure}
      \vspace{0cm}
	\resizebox{\hsize}{!}{\includegraphics{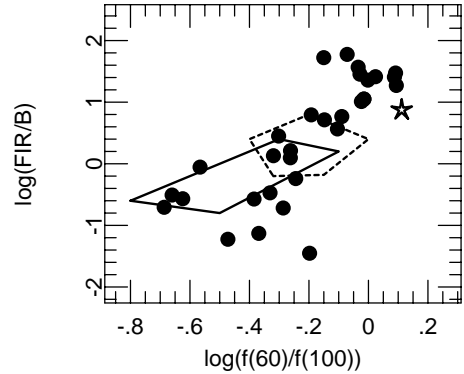}}
      \caption[]{The FIR/B luminosity ratio vs. the IRAS temperature
f$_{60}$/f$_{100}$ for the same sample as in fig. 2. The full drawn box indicates the locus of normal spiral galaxies and the hatched box the locus of blue compact galaxies according to Dultzin et al. \cite{deborah1}  }
         \label{pepa}
   \end{figure}

   \begin{figure}
      \vspace{0cm}
	\resizebox{\hsize}{!}{\includegraphics{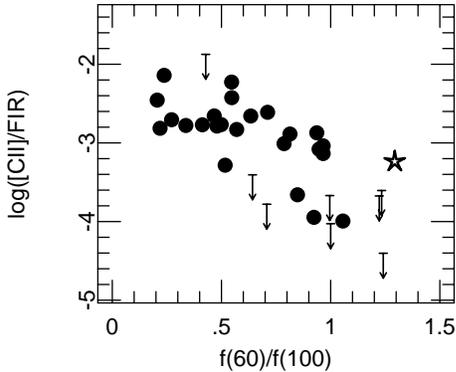}}
      \caption[]{The [CII]/FIR ratio vs. the IRAS temperature  f$_{60}$/f$_{100}$ for the same sample as in fig. 2. Upper limits are marked with arrows.}
         \label{mal2}
   \end{figure}

   \begin{figure}
	\resizebox{\hsize}{!}{\includegraphics{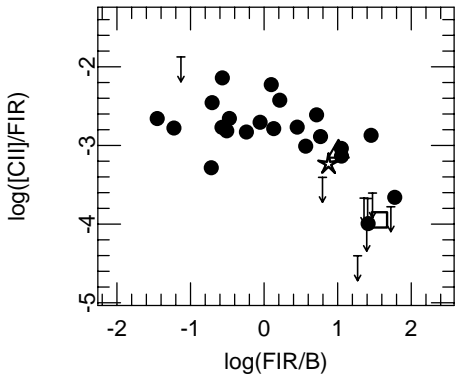}}
      \caption[]{The [CII]/FIR vs. the FIR/B luminosity ratios for the same
sample as in fig. 2. M82 is marked with a triangle and Arp220 with a square.
Upper limits are marked with arrows. }
         \label{mal1}
   \end{figure}

   \begin{figure}
      \vspace{0cm}
	\resizebox{\hsize}{!}{\includegraphics{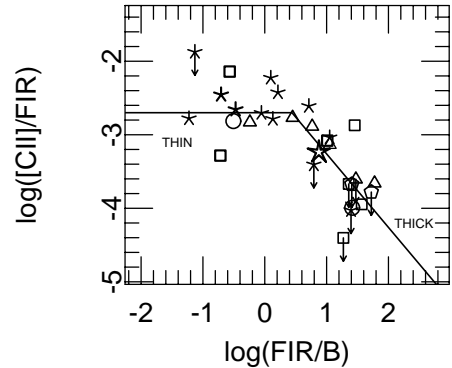}}
      \caption[]{The [CII]/FIR ratio vs. the FIR/B luminosity ratio as in fig.
5. Upper limits are marked with arrows. The different activity classes have been coded according to: Circle-Sy1, Square-Sy2, Triangle: Liner, Star: HII, Pentagon: Unclassified. The full drawn line shows an idealized approximation of the effect ot increasing optical depth with FIR/B.}
         \label{mal1b}
   \end{figure}

\section{[CII] as a probe of high redshift starbursts}

As we can see from figure 4 and 5, the [CII] line in starburst galaxies has 
a
maximum luminosity of a
few per mille of the bolometric luminosity and constitutes the peak in the
spectral distribution from
radio to X-ray. It has therefore been proposed (e.g. Petrosian et al.
\cite{petrosian}; Loeb
\cite{loeb}; Stark \cite{stark}) that it can be used as a probe in searches
for high redshift
starburst galaxies. At redshifts of z$\ge$ 5-6, where current models of 
galaxy
formation predict the
first massive galaxies to form and where optical-near IR observations start 
to
reach the confusion
limit, the [CII] line is shifted into the submillimeter (submm) region and 
could
be
detected with a suitable
submm telescope at a cold dry site like the South Pole (Stark \cite{stark}). 
The
most interesting
targets would probably be massive starburst mergers, characterized by a high
60/100 ratio and a high
FIR/B ratio. The tendencies seen in the study of Malhotra et al.
(\cite{malhotra}) however, have
created some pessimism  (e.g. Gerin \& Phillips \cite{gerin}) about the
feasibility, due to the
rapid decrease in the relative strength of the [CII] line with IRAS 
temperature.
The properties of
Haro 11 however, indicate that the conditions may be more promising than
anticipated since the
[CII]/L$_{Bol}$ of this galaxy is close to the value obtained for the normal
galaxies. If this is due
to the low metallicity of the galaxy or not remains to be explored but if so 
it
may apply on distant
metal-poor galaxies. The detection probabilities derived by Stark 
(\cite{stark})
predict that it
would be quite possible to detect a galaxy with a mass equal to that of a
present day
L$^{\star}$ galaxy at z=4-5 in a few
hours, using a 10m submm telescope at the South Pole. At the formation epoch 
a
massive galaxy
would be several magnitudes brighter than today, i.e. $\sim$ 10$^{12}$
L$_{\odot}$. Thus we would need to
merge about 10 Haro 11 type galaxies to produce this luminosity. A beam size 
of
20"-30" for a 10m
telescope corresponds to about 100 kpc at the redshift we discuss. Thus it 
would
easily encompass a
compact merging group of dwarf galaxies and make a detection feasible.
Moreover, with the advent of  planned large sub-mm/mm arrays
(e.g. ALMA), the prospects for using the [CII]158 line as a probe of high
redshift starburst would be even more interesting.

\section{Conclusions}

We have reported about ISO LWS observations of the luminous metal poor BCG 
Haro
11 in the three prominent emission lines [OI]63$\mu$, [OIII]88$\mu$ and
[CII]1588$\mu$. The galaxy is involved in an intense global starburst 
probably
due to a merger of dwarf galaxies. The ISO data and previous observations in 
the
optical-radio region in combination with model predictions are used to 
derive
information about the physical properties of gas and dust in the galaxy.

Haro 11 is one of the hottest IRAS galaxies observed and we find no trace of
cold dust. Most of the neutral hydrogen is confined to the
photodissociation regions. The main part of the gas however, appears to be 
in
ionized or
molecular state. This possibility is generally neglected when discussing the 
gas
content in starburst dwarfs, leading to serious underestimates of the
total gas mass.

We also reinvestigate the claimed correlation between [CII] emission, far-IR
luminosity and the IRAS 60/100 temperature for a number of starforming 
galaxies
of different types, active and non-active. We confirm and tighten the
correlation and note that Haro 11 deviates in the sense that its 
[CII]/L$_{FIR}$
is higher than expected. We argue that this may indicate that the previously
preferred explanation of the relationship, a decreasing efficiency with IRAS
temperature in the ejection of photoelectrons from UV-illuminated grains may 
not
be entirely correct. An alternative, or complementary explanation seems to 
be
that the optical
depth increases with increasing IRAS temperature. This would be expected if 
the
IRAS temperature correlates with compactness and/or mass of the central
starburst region. The metallicity may also play an important role, in which 
case
the somewhat pessimistic opinions about using [CII]158$\mu$ to detect high
redshift massive starburst galaxies may not be valid. The whole situation 
will
probably be clarified once a more representative sample of galaxies, 
including
both normal dwarfs and massive galaxies becomes available.

\begin{acknowledgements}

We gratefully acknowledge partial support from the
Swedish Natural Science Research Council and the Swedish Space Board.
JM and JC are supported by Spanish CICYT grant ESP98-1351. JM has been
also supported by DIGCYT grant PB93-0139. G. \"Ostlin acknowledges support
from The Swedish Foundation for International Cooperation in Research
and Higher Education (STINT).
We thank our referee for useful suggestions for improvements of the manuscript.
This research has
made use of the NASA/IPAC
Extragalactic Database (NED)
which is operated by the Jet Propulsion Laboratory, California Institute
of Technology, under contract with the National Aeronautics and Space
Administration.

\end{acknowledgements}

\end{document}